\documentclass[prl,twocolumn,amsmath,amssymb,showpacs,superscriptaddress,floatfix]{revtex4}
\def\vc{\mathbf}
\newcommand{\celc}[0]{^\circ\mathrm{C}}
\newcommand{\kT}[0]{k_\mathrm{B}T}

\usepackage{graphicx}
\usepackage{float}
\usepackage{color}
\usepackage{soul}
\usepackage{epsfig}
\usepackage{textcomp}
\begin{document}
\title{Phonons in two-dimensional soft colloidal crystals}
\author{Ke Chen }
\affiliation{Department of Physics and Astronomy, University of Pennsylvania, Philadelphia, Pennsylvania 19104, USA}
\affiliation{Beijing National Laboratory for Condensed Matter Physics and Key Laboratory of Soft Matter Physics, Institute of Physics, Chinese Academy of Sciences, Beijing 100190, China}
\author{Tim Still }
\affiliation{Department of Physics and Astronomy, University of Pennsylvania, Philadelphia, Pennsylvania 19104, USA}
\author{Samuel Schoenholz }
\affiliation{Department of Physics and Astronomy, University of Pennsylvania, Philadelphia, Pennsylvania 19104, USA}
\author{Kevin B. Aptowicz }
\affiliation{Department of Physics, West Chester University, West Chester, Pennsylvania, 19383, USA}
\author{Michael Schindler}
\affiliation{Laboratoire PCT, Gulliver CNRS-ESPCI UMR 7083, 10 rue Vauquelin,75231 Paris Cedex 05, France}
\author{A. C. Maggs}
\affiliation{Laboratoire PCT, Gulliver CNRS-ESPCI UMR 7083, 10 rue Vauquelin,75231 Paris Cedex 05, France}
\author{Andrea J. Liu}
\affiliation{Department of Physics and Astronomy, University of Pennsylvania, Philadelphia, Pennsylvania 19104, USA}
\author{A. G. Yodh}
\affiliation{Department of Physics and Astronomy, University of Pennsylvania, Philadelphia, Pennsylvania 19104, USA}
\date{\today}
\begin{abstract}
The vibrational modes of pristine and polycrystalline monolayer colloidal crystals composed of thermosensitive microgel particles are measured using video microscopy and covariance matrix analysis. At low frequencies, the Debye relation for two dimensional harmonic crystals is observed in both crystal types; at higher frequencies, evidence for van Hove singularities in the phonon density of states is significantly smeared out by experimental noise and measurement statistics. The effects of these errors are analyzed using numerical simulations. We introduce methods to correct for these limitations, which can be applied to disordered systems as well as crystalline ones, and we show that application of the error correction procedure to the experimental data leads to more pronounced van Hove singularities in the pristine crystal. Finally, quasi-localized low-frequency modes in polycrystalline two-dimensional colloidal crystals are identified and demonstrated to correlate with structural defects such as dislocations, suggesting that quasi-localized low-frequency phonon modes may be used to identify local regions vulnerable to rearrangements in crystalline as well as amorphous solids.
\end{abstract}
\maketitle
\section{Introduction}
Recently, video microscopy has been cleverly employed to extract information about the dynamical matrix of ordered~\cite{keim_prl_2004, reinke_prl_2007} and disordered~\cite{ghosh2010, klix_prl_2012} colloidal systems from particle position fluctuations. This technical advance has opened a novel experimental link between thermal colloids and traditional atomic and molecular materials~\cite{baumgartl_prl2007, kaya2010,chen2010,yunker_prl2011,green_pre2011,chen2011, ghosh2011, ghosh2011_2,yunker_pre2011,tan_prl2012}. Along these lines, one ubiquitous feature of atomic glasses, the so-called ``boson peak" due to an excess number of vibrational modes at low frequency~\cite{phillipsbook, pohl02}, has been observed in disordered colloidal packings \cite{chen2010, ghosh2010,kaya2010}. Furthermore, connections have been established between the ``soft spots" associated with quasi-localized low-frequency vibrational modes and localized particle rearrangements in disordered colloids \cite{cooper_np2008, xu_epl2010, manning2011, chen2011, ghosh2011}, reinforcing the possibility that such phenomena might exist in atomic and molecular glasses, as well. 

The present paper has two primary themes. First, we use nearly perfect crystals to measure the phonon density of states.  To recover van Hove singularities, arguably the most prominent feature of phonons in crystals, we introduce error correction procedures which are applicable even to disordered systems and which reduce the amount of data needed to reliably perform such analyses by orders of magnitude. Second, we study {\it imperfect} crystals in order to probe directly the effects of defects on phonon modes. We find that structural defects in the {\it imperfect} two-dimensional colloidal crystals are spatially correlated with quasi-localized low-frequency phonon modes. Thus, our experiments extend ideas about quasi-localized low-frequency modes and flow defects in colloidal glasses~\cite{manning2011, chen2011, ghosh2011} to the realm of colloidal crystals and suggest that phonon properties can be used to identify crystal defects which participate in the material's response to mechanical stress.

Specifically, we employ video microscopy and covariance matrix analysis to explore the phonons of various two-dimensional soft colloidal crystals. By studying two-dimensional crystals~\cite{peeters_pra_1987}, we avoid significant complications~\cite{schindler_2012, lemar_epl2012} encountered by previous experiments which analyzed two-dimensional image \textit{slices} within three-dimensional colloidal crystals to derive phonon properties~\cite{kaya2010, ghosh2011_2}. Our work is also complementary to earlier experiments by Keim \textit{et al.}~\cite{keim_prl_2004} and Reinke \textit{et al.}~\cite{reinke_prl_2007} which studied colloidal crystals stabilized by \textit{long-ranged} repulsions and found good quantitative agreement between the dispersion relation measured from particle fluctuations and theoretical expectation. By contrast, the present experiments measure not only the dispersion relations for crystals of particles with \textit{short-range} interactions, but also the density of vibrational states, which turns out to be far more sensitive to statistical error. In addition, we identify a narrow band of modes in imperfect crystals that are associated with crystal defects.

\section{Experiment}
The experiments employed poly($N$-iso\-propyl\-acryl\-amide) or PNIPAM microgel particles, whose diameters decrease with increasing temperature. Particle diameters are measured to be 1.4 \textmu m at 22 $\celc$ by dynamic light scattering; the sample polydispersity was also determined by light scattering to be approximately 5\%. PNIPAM particles are loaded between two coverslips, and crystalline regions are formed as the suspension is sheared by capillary forces. The samples are then hermetically sealed using optical glue (Norland 65) and thermally cycled between 28 $\celc$ and room temperature for at least 24 hours to anneal away small defects. Particle softness permits the spheres to pack densely with stable contacts and yet still exhibit measurable thermal motions. At room temperature, the particles are immobile due to swelling. The samples are then slowly heated from room temperature until noticeable motion is observed at 24.6 $\celc$. Before data acquisition, samples equilibrate for 4 h on the microscope stage. Bright-field microscopy images are acquired at 60 frames per second with a total number of frames of 40,000. An image shutter speed of 1/4000 second is used. Care was taken so that no structural rearrangements occurred during the experimental time window. Each image contains about 3000 particles in the field of view. The trajectory of each particle in the video was then extracted using standard particle-tracking techniques~\cite{crocker_1996}. The particle position resolution is approximately 6 nm. Crystal quality is characterized by Fourier transformation of the microscopy images, and by spatial correlations of the bond orientational order parameter, $\Psi_{6}$~\cite{ suppl}. Particle spacing is measured to be 1.19 \textmu m, indicating a small compression of PNIPAM spheres ($\approx70$~nm compared to the hydrodynamic diameter).

To obtain intrinsic vibrational modes of the colloidal crystal samples, we employ covariance matrix analysis. Specifically, a displacement vector $\vc{u}(t)$ that contains the displacement components of all particles from their equilibrium positions is extracted for each frame. A covariance matrix $C$ is constructed with
$C_{ij}= \langle u_i u_j\rangle_{t}$, where $i$ and $j$ run through all particles and coordination directions. To quadratic order, the stiffness matrix $K$, which contains the effective spring constants between particles, is proportional to the inverse of $C$ with $K=\kT C^{-1}$. 
Thus, by measuring the relative displacement of particles, interactions between particle pairs can be extracted and the dynamical matrix can be constructed, i.e.,
\begin{equation}
D=\frac{K}{m}=\frac{\kT(C^{-1})}{m}~,
\end{equation}
where $m$ is the particle mass. The dynamical matrix yields the eigenfrequencies and eigenvectors of the \emph{shadow system}: the system of particles with the exact same interactions and geometry as the colloidal particles, but without damping. This approach of measuring phonons permits direct comparison to theoretical models, e.g., models that might be used to understand atomic and molecular crystalline solids.

\begin{figure}[t]
\includegraphics[width= 9 cm]{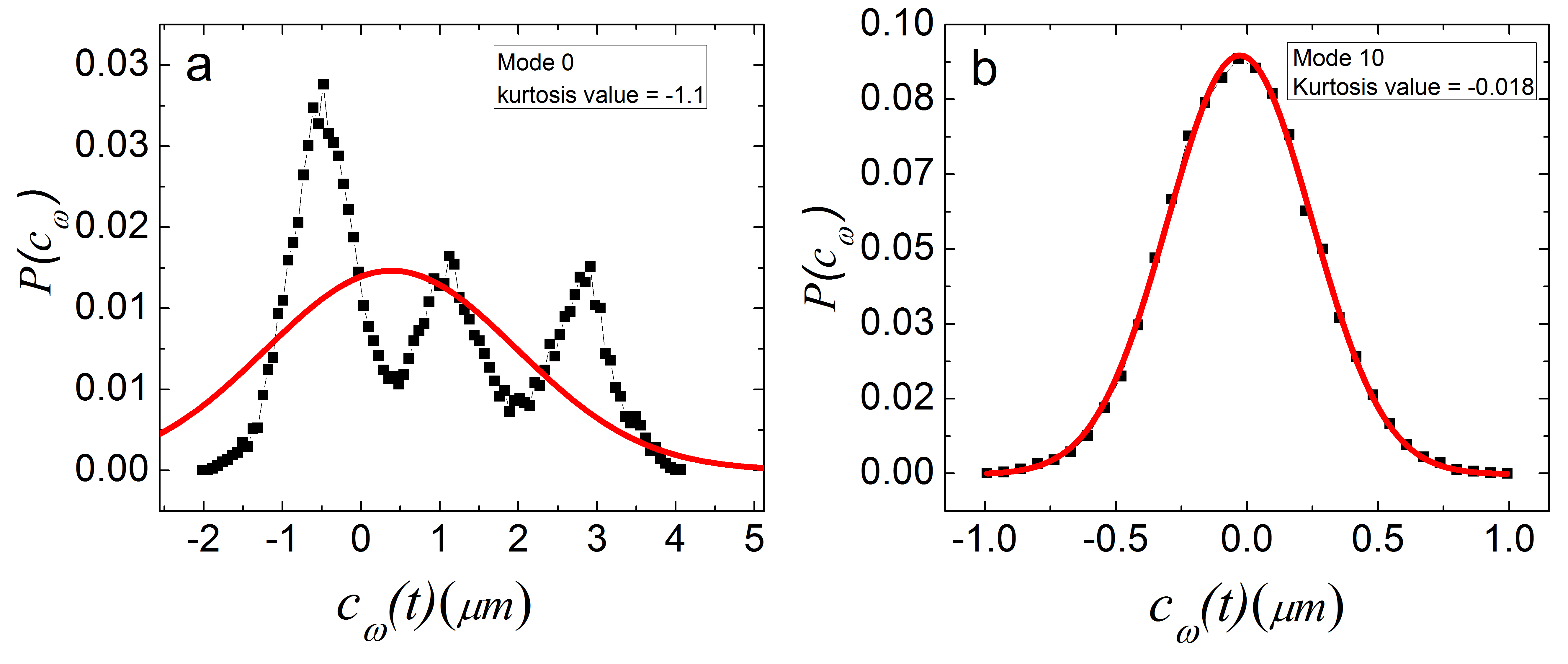}
\caption{\textcolor[rgb]{0.98,0.00,0.00}{(color online)} Distribution of instantaneous projection coefficient $c_\omega(t)$ for (a) the lowest-frequency mode, mode 0; and (b) the tenth lowest-frequency mode. Red lines are Gaussian fits to the distributions. }  
\label{fig:kurt}
\end{figure}

\begin{figure*}[t]
\includegraphics[width=18 cm]{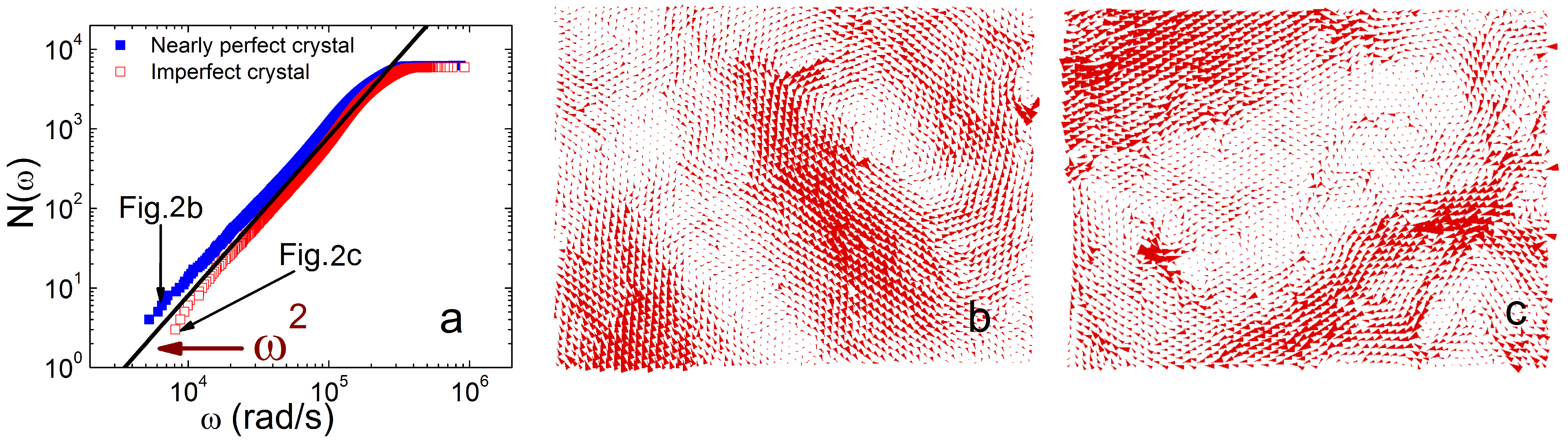}
\caption{\textcolor[rgb]{0.98,0.00,0.00}{(color online)} Phonon modes in 2D colloidal crystals.~(a) Accumulated mode number, $N(\omega)$, as a function of frequency, for a nearly perfect crystal (blue squares) and an imperfect one (red square); $\omega^2$ is drawn for comparison (black line); small arrows point to the modes whose real space vector distributions are plotted in panels b and c.~(b, c) Spatial distribution of a low-frequency mode for the (b) nearly perfect and (c) imperfect crystal; the direction and magnitude of polarization vectors are represented by the direction and size of the arrows.} 
\label{fig:sgl}
\end{figure*}

Displacement covariance analysis assumes that the local curvature of the multi-dimensional potential energy landscape of the system $V(\vc{u}_{1},\vc{u}_{2},...,\vc{u}_{N})$  is harmonic near the equilibrium configuration. However, this assumption may not be satisfied for some modes obtained from experiment. To test the harmonicity of individual modes,  we define an instantanous projection coefficient $c_{\omega}(t)= \vc{u}(t)\cdot \vc{e}_{\omega}$, where $\vc{e}_{\omega}$ is the eigenvector with frequency $\omega$. Consider the instantanous potential energy of the system 
\begin{eqnarray}
V(\vc{u(t)}) &=& \frac{1}{2}\vc{u(t)} K \vc{u(t)}^* \propto (\sum\limits_{\omega} c_{\omega}(t) \vc{e}_{\omega}) K (\sum\limits_{\omega} c_{\omega}(t) \vc{e}_{\omega}^*)  \nonumber \\
                    &\propto& \sum\limits_{\omega} (\omega c_{\omega}(t))^2.
\end{eqnarray}

The contribution to the system potential energy from mode $\omega$ is $E_{\omega}(t) \propto \omega^2 c_{\omega}(t)^2$. This potential energy component is the result of thermal fluctuations, and therefore it should be characterized by the Boltzman distribution, i.e., 
\begin{equation}
P(E_{\omega}(t)) \propto e^{-\frac{\omega^2 c_{\omega}(t)^2}{2kT}}.
\end{equation}

A Gaussian distribution of the instantaneous projection coefficient indicates that the potential energy component from mode $\omega$ increases parabolically with displacement, i.e., the curvature of the system potential along the direction of that particular eigenvector is harmonic. The deviation from a Gaussian distribution is typically quantified by the kurtosis value.  In our experiments, we find that a few of the lowest frequency modes, typically fewer than five modes, have kurtosis values larger than 0.2 and can even display a multi-modal distribution of $c_{\omega}$ as shown in Fig.~\ref{fig:kurt}(a). Most modes have kurtosis values less than 0.1, as plotted in Fig.~\ref{fig:kurt}(b). The harmonic assumption is not valid for modes with high kurtosis values, so we have excluded modes with kurtosis values greater than 0.2 from our analysis. The high kurtosis values for those modes may result from several factors, including undersampling of the lowest energy basins~\cite{hess_pre2000} and tiny shifts of equilibrium positions during experiment~\cite{henkes_sm2012}. 

 The local spring constants are measured to be uniformly distributed in space.  Centrifugal compression experiments~\cite{nordstrom_2010} show for PNIPAM, the inter-particle interaction potential is consistent with the Hertzian form. Vertical fluctuations are primarily due to particle polydispersity and are small; their effects on the modes obtained by the covariance method are calculated to be negligible (i.e., less than 4\%)~\cite{chen2010}. 

\section{Phonon Density of states in colloidal crystals}
At low frequencies, Debye scaling requires that the phonon density of states $D(\omega)$ scales with $\omega^{d-1}$, where $d$ is the dimensionality of the system, or that $N(\omega)$, the number of modes with frequencies below $\omega$, scales as $\omega^d$~\cite{merminbook}. Debye scaling for two-dimensional crystals is clearly exhibited by both our perfect and imperfect monolayer colloidal crystals (see Fig.~\ref{fig:sgl}(a)). For more than one decade, $N(\omega)$ follows a power law close to 2, as expected. The lowest frequency modes typically exhibit wave-like features as shown in Figs.~\ref{fig:sgl}(b) and (c) (more real space vector plots of low frequency modes can be found in supplementary material~\cite{suppl}). Thus we conclude that the Debye scaling observed in Fig.~\ref{fig:sgl}(a) is due to wave-like ``acoustic modes." 
\begin{figure}[t]
\includegraphics[width=8 cm]{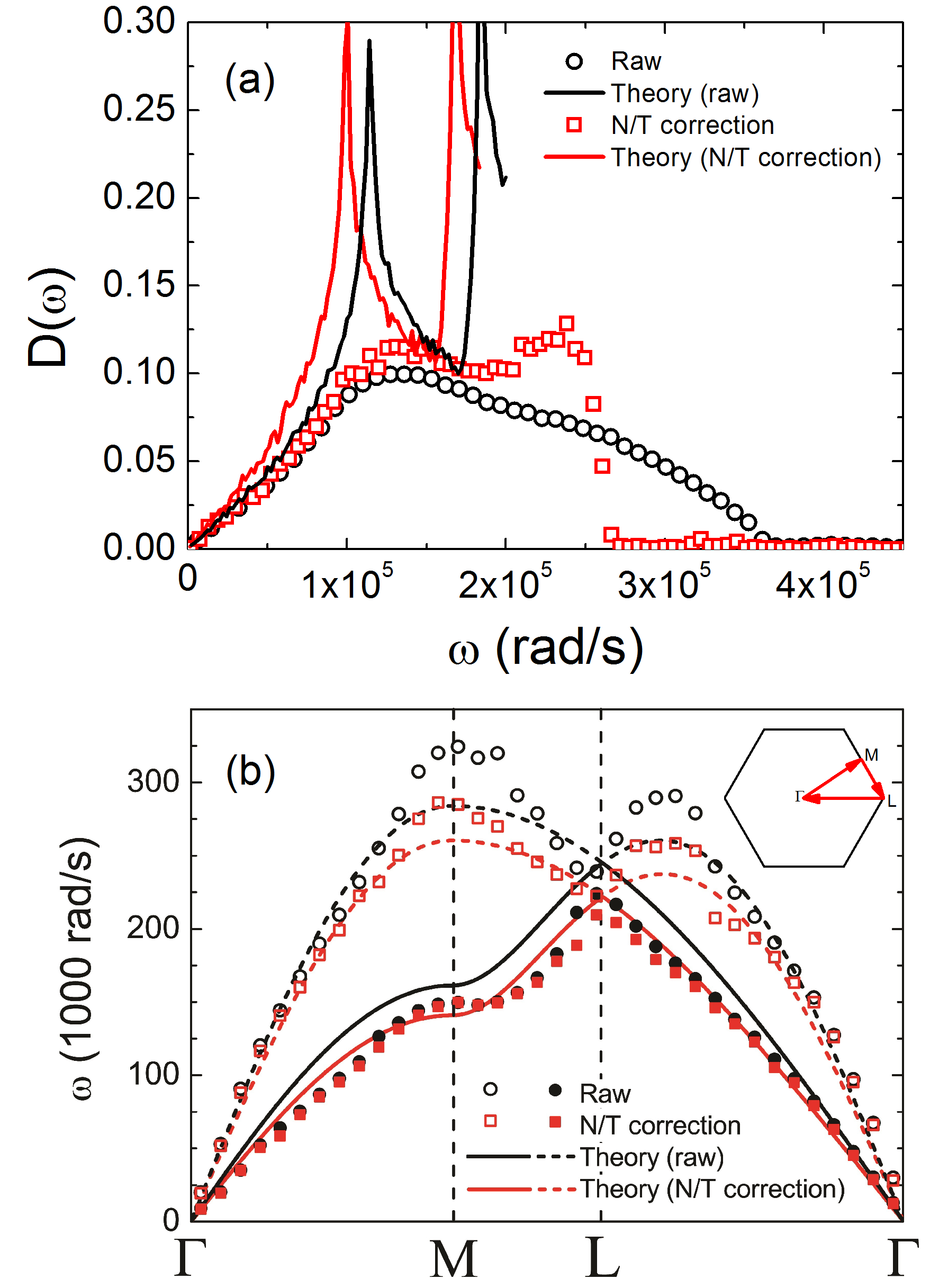}
\caption{\textcolor[rgb]{0.98,0.00,0.00}{(color online)}(a) Phonon density of states of two dimensional colloidal crystals. The DOS is obtained from experimental data (black circles), after N/T extrapolation(red squares). The numerically generated DOS for the harmonic triangular lattice (with matched sound speeds) is plotted as a guide to the eye (solid lines). (b) Dispersion curves for longitudinal (open symbols) and transverse (filled symbols) modes along high-symmetry directions, including uncorrected experimental data (black circles), and data after N/T extrapolation (red squares). Theoretical expectations are plotted in dashed (longitudinal) and solid (transverse) lines with matched colors. Inset: high symmetry directions in reciprocal space.} 
\label{fig:dos}
\end{figure}

At higher frequencies, deviations from Debye behavior are observed in the phonon density of states, $D(\omega)$, as shown in Fig.~\ref{fig:dos}(a) for the pristine crystal. For triangular two-dimensional crystals with harmonic interactions, the density of states has two van Hove singularities, one for longitudinal modes and one for transverse modes (solid lines in Fig.~\ref{fig:dos}(a)); these singularities are expected to arise at the boundary of the first Brillouin zone. In contrast, Fig.~\ref{fig:dos}(a) shows that the experimentally measured $D(\omega)$ (black open circles) exhibits a smooth peak at an intermediate frequency and a shallow shoulder at a higher frequency.

\section{Error analysis and corrections}
We identify the peak and shoulder as vestiges of van Hove
singularities. Several factors may contribute to the rounding of van
Hove singularities in a colloidal crystal. For example, particle
polydispersity may break translational symmetry for the largest wave
vectors, wherein van Hove singularities appear. Further, the statistics
associated with the finite number of frames (i.e., the finite number of
temporal measurements), as well as uncertainties in locating particle
positions, can introduce noise into the covariance matrix and thus
into its eigenvalues and eigenvectors \cite{henkes_sm2012}. In the following, we discuss
these effects, and we show how to recover some of the expected
behavior by applying corrections to the experimental data.

We first show that in the limit of perfect statistics (i.e., wherein
the covariance matrix is calculated from an infinite number of time
frames), measurement error modifies the effective interactions between
particles but does not smooth out the peaks. We do this by first
studying the distribution of our original system, and then showing
that the distribution of a system which has been visualized with
finite resolution, $\sigma$, is described by an effective potential
energy which we calculate as an expansion in $\sigma$.

Consider a colloidal system with particles interacting with potential
energy $V(\{r_i \})$, where $i$ labels the $2N$ coordinates of the
particles. The  distribution of states at equilibrium is
\begin{equation}
  \rho(\{r_i\}) = e^{-\beta V(\{r_i\})}/Z.
\end{equation}
with $\beta=1/k_BT$, and $Z$ normalizes the distribution.  We now imagine that the system is
observed with independent Gaussian errors $\xi_i$ on the positions of
each particle. In this case, we will observe an effective distribution
for the degrees of freedom which is a convolution of the true position
with the observation error:
\begin{equation}
  \rho(\{r_i'\})= \frac{1}{Z} \int  e^{- \beta V(\{r_i\}) } \prod_i \delta (r_i'-r_i - \xi_i)
  \frac{ e^{-\xi_i^2/2\sigma^2} }{\sqrt{2\pi} \sigma } \;  dr_i, \nonumber
\end{equation} 
where $\sigma$ is the width of the distribution of $\xi$.  The
integral over $r$ gives an effective weight in the exponential,
\begin{equation}
  A = \beta V( \{ r'_i - \xi_i \}) + \sum_i \xi_i^2/2 \sigma^2,
\end{equation}
and expanding this potential (assumed smooth) to second order
gives:
\begin{equation}
  A = \beta V(\{r_i'\}) + \beta \sum_i \xi_i V_i + \beta \sum_{ij} \frac{\xi_i \xi_j V_{ij}}{2} + \sum_i
  \frac{\xi_i^2  }{2 \sigma^2}+\dots
\end{equation}
where $V_{i}$ is the derivative of the potential with respect to the
$i$th coordinate. Since the variables $\xi_i$ are unknown we integrate
over them to find the distribution of the observed
variables; we expand the result keeping terms up to second order in
the resolution $\sigma$.
\begin{equation}
  A= \beta V -   \sum_i \frac{ \beta^2 \sigma^2 V_i^2}{2}  + \frac{\beta
    \sigma^2}{2 }   \sum_i
  V_{ii}  .
  \label{eq:action}
\end{equation}
This is our main result for the effective potential of a system
observed with a finite resolution.

The ratio of the coefficients of the two terms is fixed due to the
fact that the integral of the probability should not change, Thus
\begin{equation}
\int e^{-A} \prod_i dr_i'\approx \int e^{-\beta V}  ( 1 +\sum_i { \frac{ \beta^2 \sigma^2 V_i^2}{2} } - {\frac{\beta
\sigma^2}{2 }
 } \sum_i V_{ii}
)\label{eq:parts}
\end{equation}
Notice, if we integrate by parts, then $V_{ii}$ in Eq.~(\ref{eq:parts}) cancels the contribution in
$V_i^2$, and the probability distribution remains normalized.

Let us specialize to the case of harmonic nearest neighbor
interactions. Then $V_{ii}$ is no more than a shift in the zero of the
energy in the Hamiltonian. The contribution $V_i^2$ however is a
quadratic contribution which introduces new interactions out to second
nearest neighbours in the system.  This effective interaction will
shift the van Hove singularities, but cannot smooth them out. It is
interesting to note that mathematically similar perturbation series
can be found in quantum mechanics in the Wigner semiclassical
expansion~\cite{wigner} as well as in the analysis of integration
errors in the leap-frog integrator~\cite{warnner_book}. In quantum
mechanics the role of $\sigma$ is replaced by the thermal wavelength
$\lambda \sim \hbar\sqrt{\beta/m}$, which also renders the position of
the particles uncertain.

\begin{figure}[t]
\includegraphics[width=8 cm]{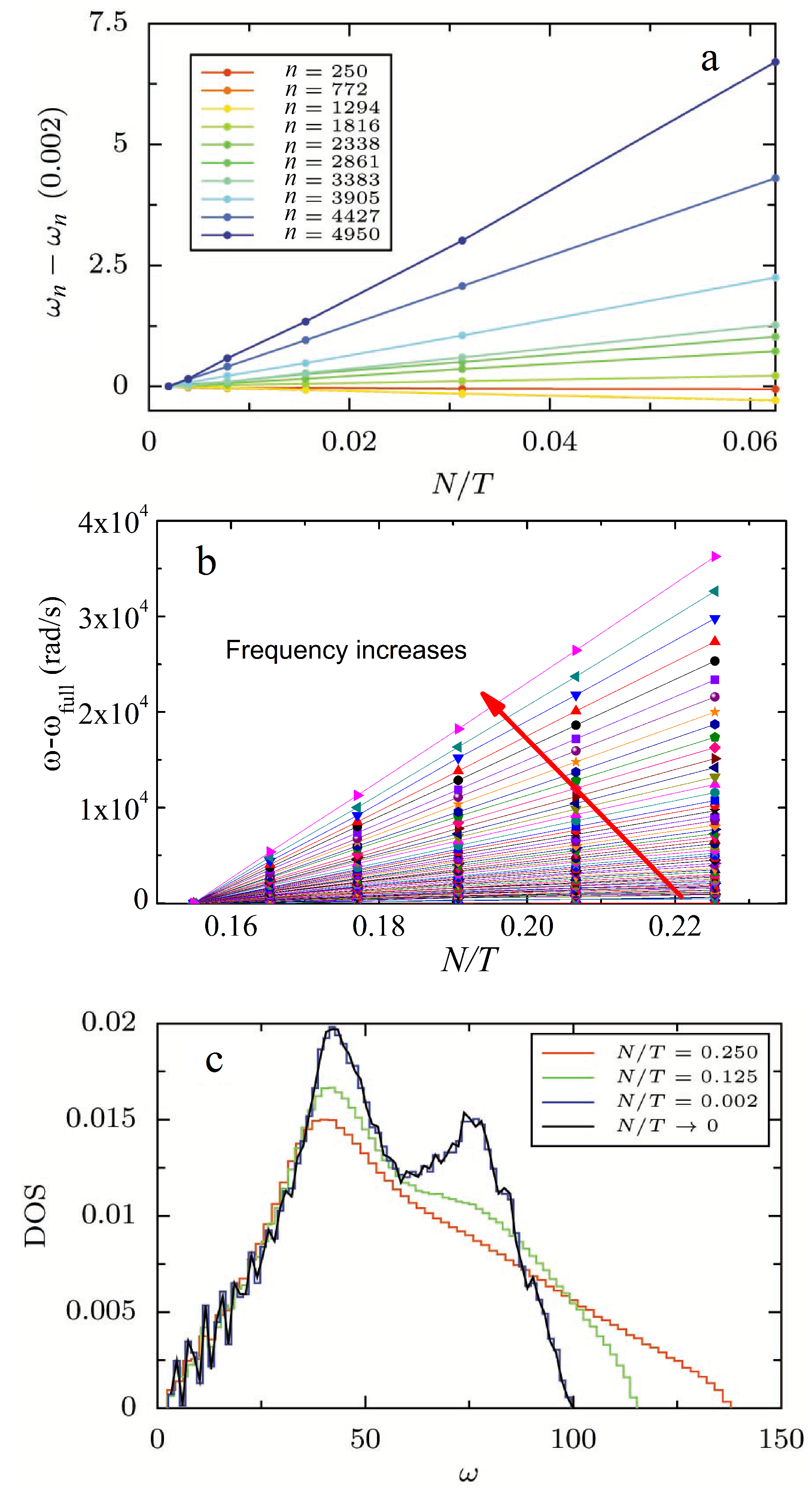}
\caption{\textcolor[rgb]{0.98,0.00,0.00}{(color online)} Correction for a finite number of observation frames. (a). Linear dependence of eigenfrequencies on $N/T$ from simulation. $n$ is the mode index which increases from low to high frequencies. For better visualization, a constant, the corresponding mode frequency for $R = 0.002$, has been subtracted for each curve. (b). Linear convergence of eigenfrequencies from experimental data with $N/T$. The vertical axis plots the frequencies minus the frequency from the full length of the video, $\omega_{full}$. (c). Density of states for different ratios of $N/T$ from simulation.} 
\label{fig:extra}
\end{figure}

To understand the smearing/rounding of the van Hove singularities, we next consider the opposite case, i.e., wherein no measurement error exists but there exists an error associated with the quality of the statistics used to calculate the covariance matrix. A key quantity for this analysis is the parameter, $R=N/T$, where $N$ is the number of degrees of freedom in the sample, and $T$ is the number of independent time frames (observation frames) used in construction of the covariance matrix. $R < 1$ ensures that the covariance matrix is constructed from independent measurements while $R \rightarrow 0$ corresponds to the limit of perfect statistics. In our experiment, $R \gtrsim  0.15$. 

For nonzero values of $R$, the noise in the matrix gives rise to a systematic error in the density of states. In particular, it smoothes the van-Hove peaks and shifts the top of the spectrum to higher $\omega$. Random matrix theory suggests that the eigenvalues distribution should converge linearly to its limiting $R = 0$ values in disordered systems~\cite{burda_pra2004}. To study this effect, we performed molecular dynamics simulations of crystalline samples and calculated eigenfrequencies from the constructed displacement covariance matrix. The eigenfrequencies indeed converge linearly with $R$ to the values at perfect statistics ($R = 0$), as plotted in Fig.\ref{fig:extra}(a). A similar convergence of eigenfrequencies is also observed in our experimental data, when the experiment video is truncated into different lengths, as shown in Fig. \ref{fig:extra}(b). The comparison between the raw data (black circles) and extrapolated data (red squares) in Fig.~\ref{fig:dos}(a) shows that the corrections from extrapolation are larger at higher frequencies, as expected, and that the extrapolation converts the shoulder at the second van Hove singularity into a small peak.  The theoretical curves are different in the two cases because extrapolation affects the speed of sound, which we use to fit to theory.

Our numerical simulations of crystalline particle packings also show that the density of states obtained from extrapolating to $R \rightarrow 0$ agrees to within noise with the result for $R = 0.002$ as plotted in Fig.\ref{fig:extra}c.  Note that $R = 0.002$ corresponds to nearly two orders of magnitude more data than are available experimentally ($R \gtrsim 0.15$).  Thus, our simulations indicate that linear extrapolation to $R \rightarrow 0$ makes the covariance matrix technique far more powerful in practice.

\begin{figure}[H]
\includegraphics[width=\linewidth]{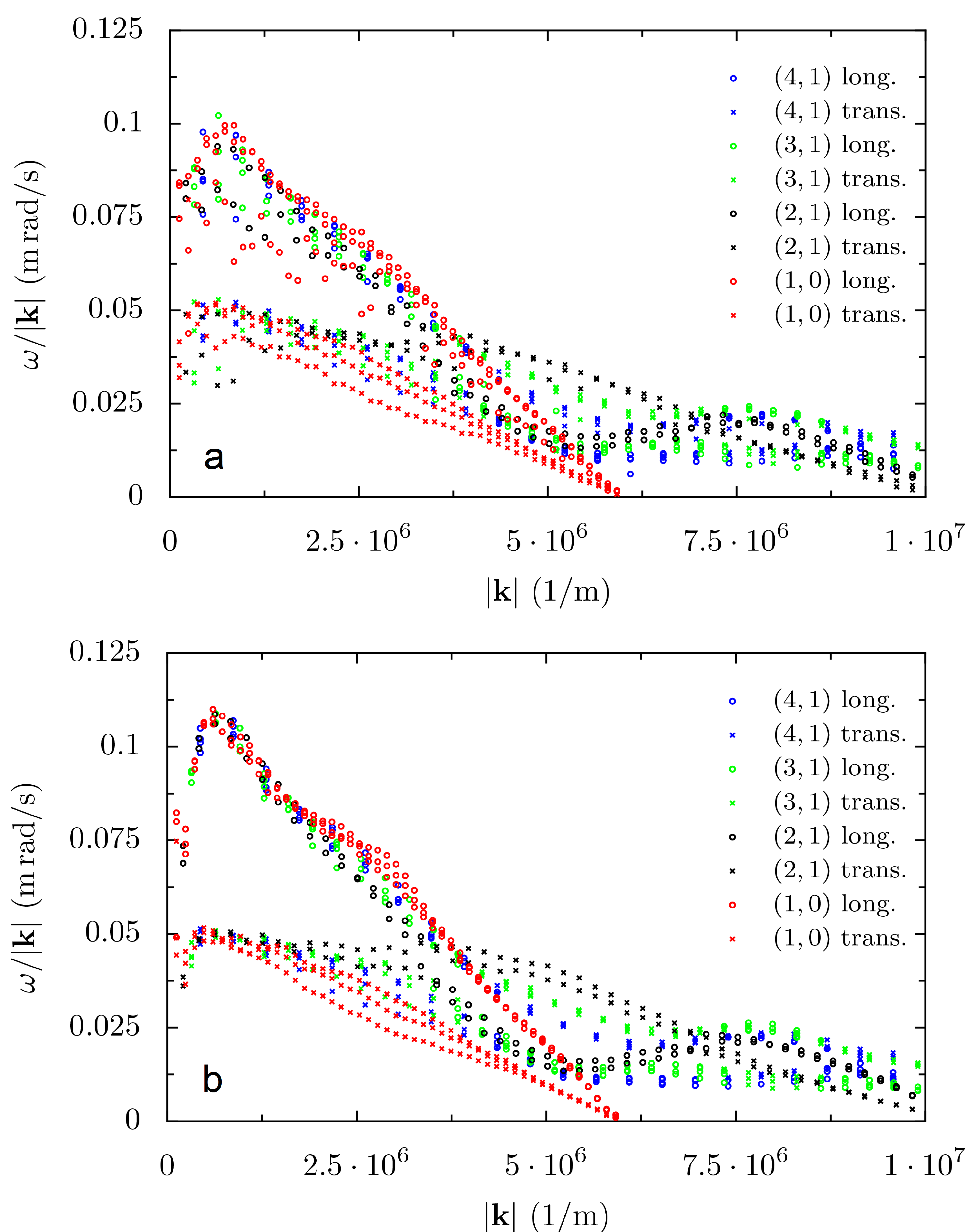}
\caption{\textcolor[rgb]{0.98,0.00,0.00}{(color online)} Correction for limited field of veiw. (a). Dispersion curves along different directions derived from uncorrected data. Circles represent longitudinal modes and crosses represent transverse modes. Directions are indexed using the basis vectors of the reciprocal lattice.  (b). Dispersion curves along different directions after the Hann window function correction. } 
\label{fig:Hann}
\end{figure}

\begin{figure*}[t]
\includegraphics[width=18cm]{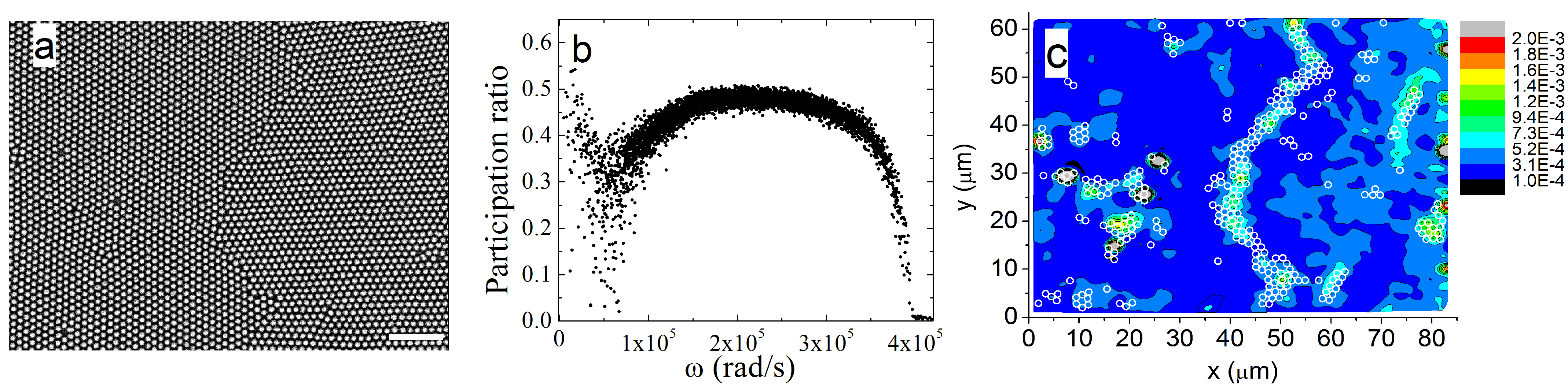}
\caption{\textcolor[rgb]{0.98,0.00,0.00}{(color online)} Low-frequency modes in a colloidal crystal with defects.~(a). A snapshot of an imperfect PNIPAM crystal with a grain boundary in the middle of the field of view; the scale bar is 10 \textmu m. (b). Participation ratio for eigenmodes in crystal with defects. (c). Color contour plots indicating polarization magnitudes for each particle, summed over the low-frequency modes with participation ratios less than 0.2. Circles indicate ``defect" particles identified by local structural parameters.} 
\label{fig:gb}
\end{figure*}

We also studied the effect of errors on the dispersion relation. For each eigenmode obtained from the covariance matrix, Fourier transformation of the longitudinal and transverse components of the eigenvector yields two spectral functions, $f_L$ and $f_T$, respectively~\cite{silbert09, vitelli10, still13arxiv}.
\begin{equation}
f_T(k,\omega)=\left\langle \left| \sum_n \widehat{\mathbf{k}}\times\mathbf{e}_{\omega,i}\exp(i\mathbf{k}\cdot\mathbf{r}_i)\right|^2\right\rangle,
\label{eq:ft}
\end{equation}
\begin{equation}
f_L(k,\omega)=\left\langle \left| \sum_n \widehat{\mathbf{k}}\cdot\mathbf{e}_{\omega,i}\exp(i\mathbf{k}\cdot\mathbf{r}_i)\right|^2\right\rangle, 	
\label{eq:fl}
\end{equation}
where $\mathbf{e}_{\omega,i}$ is the polarization vector on particle $i$ in mode $\omega$, $\mathbf{k}$ is the wavevector, $\mathbf{r}_n$ is the equilibrium position of each particle, and the brackets indicate an average of directions  $\widehat{\mathbf{k}}$  (for crystals: identical crystallographic directions).
We calculated these spectral functions of the eigenmodes from using data from an experimental colloidal crystal along high-symmetry directions; for each phonon frequency, the wavevector corresponding to the maximum of the spectral function, $k_{max}(\omega)$, is readily extracted and the dispersion relation is thus determined. Note this method does not require any underlying periodicity, and it was recently applied to colloidal glasses~\cite{still13arxiv}.  The binned dispersion curves [black symbols in Fig.~\ref{fig:dos}(b)] largely follow the theoretical expectation, obtained by fitting to the low-frequency part of the curve to obtain the longitudinal and transverse speeds of sound, as shown in Fig.~\ref{fig:dos}(b). However, the measured dispersion relation has a stronger dependence on $k$, especially for the longitudinal branch. We extrapolated to the limit of perfect statistics, where $R=N/T \rightarrow 0$. We find from simulations that, like the mode frequencies, the dispersion relation approaches its limiting $R=0$ value linearly in $R$. Extrapolation of the data [red symbols in Fig.~\ref{fig:dos}(b)] leads to excellent agreement with theory. Thus, \emph{the dispersion relation appears far less sensitive than the density of vibrational states to both of the leading sources of error} in the covariance matrix technique, i.e., less sensitive to measurement errors in the positions of the particles and due to limited statistics.

By extrapolating $R$ to zero, we obtain the bulk modulus of  $B=2.4\times10^{-5}$~Pa~m ($\pm8\times10^{-6}$), and the shear modulus of $G=5.4\times10^{-6}$~Pa~m ($\pm1.8\times10^{-6}$). The measured shear modulus is in line with bulk rheology measurements of PNIPAM suspensions~\cite{senff_jcp1999}.

The finite spatial observation window in the experiment also introduces errors into the obtained spectrum. For a crystal, the dispersion relation can be derived from the covariance matrix in reciprocal space. Specifically, for a given wave vector $\mathbf{k}$, a $2\times2$ matrix can be constructed as $C_{ij}(\mathbf{k}, -\mathbf{k}) = \langle u_{i}(\mathbf{k})u_{j}(\mathbf{k})) \rangle$, where $u_{i}(\mathbf{k}) = \sum_re^{i\mathbf{k}\cdot\mathbf{r}}u_{i}(\mathbf{r})$, summed over an infinite number of particles in a crystal lattice. In experiments, the observation window is limited, and $u_{i}(\mathbf{k})$ is the result of an integral a over finite number of particles; thus, the dispersion curves obtained suffer from truncation errors. This error can be ameliorated by applying the Hann window function to the Fourier transformation as proposed in Ref.\cite{schindler_2012}. The Hann window function correction significantly improves the obtained dispersion curves, particularly for the longitudinal branch, as shown in Fig.\ref{fig:Hann}. We note that the Hann window correction applies only to crystalline samples with nearly perfect lattices, and can not be used in glasses or crystalline samples with significant numbers of defects.

\section{Soft modes in imperfect crystals}
Finally, we explored phonons in the imperfect two-dimensional colloidal crystal shown in Fig.~\ref{fig:gb}(a). Most of the low frequency modes are extended and wave-like [as in Fig.~\ref{fig:sgl}(c)], consistent with the observation of Debye scaling in the accumulated number of modes, $N(\omega)$, shown in Fig.~\ref{fig:sgl}(a). To take a closer look at the nature of the phonon modes in the imperfect crystal, we calculated the participation ratio, $p(\omega)$, which measures the degree
of spatial localization of a mode $\omega$ and is defined as $p(\omega) = \left(\sum_i |\mathbf{e}_{\omega,i}|^2\right)^2/(N\sum_i |\mathbf{e}_{\omega,i}|^4)$, where $\mathbf{e}_{\omega,i}$ is the polarization vector on particle $i$ in mode $\omega$. A participation ratio of 1 indicates collective translational motion; a perfect plane wave has a participation ratio of 1/2, and a mode localized on one particle has a participation ratio of $1/N$, where $N$ is the total number of particles in the system~\cite{bell_jpc1970}. Fig.~\ref{fig:gb}(b) shows that while most of the modes have participation ratios near $0.5$, close to the value expected for a plane wave, a few of the low-frequency modes have a significantly smaller participation ratio.

The quasi-localized low-frequency modes observed in Fig.~\ref{fig:gb}b are reminiscent of those characteristically observed in glasses~\cite{chen2010,chen2011}. In jammed packings, such modes have unusually low barriers to rearrangements~\cite{xu_epl2010} and have been used to identify regions that serve as flow defects when the packings are sheared~\cite{manning2011} or dilated~\cite{chen2011}. In crystalline systems, it is known that rearrangements tend to occur at crystal defects, particularly dislocations. Our observation of quasi-localized modes in imperfect crystals therefore raises the question of whether the modes are spatially concentrated near structural defects such as dislocations.

In the colored contour map in Fig.~\ref{fig:gb}c, we plot the spatial distribution of the quasi-localized low-frequency modes with a participation ratio less than 0.2, i.e., $\frac{1}{N_{0.2}}\sum\limits_{pr(\omega)<0.2}({\mathbf{e}_{\omega,i})^2}$. Some particles contribute significantly to more than one soft mode, which results in regions with lighter colors. The white circles in Fig.~\ref{fig:gb} indicate structural defects in the crystal sample, identified by local structural parameters. Here a particle is identified as a ``defect" particle if the number of its nearest neighbors is not 6, \textit{or} the magnitude of the local bond orientational order parameter $\Psi_{6}$ =$\frac{1}{N_{nn}}$$\sum_{k}^{N_{nn}}{e^{6i\theta_{jk}}}$ is less than 0.95.

The spatial correlation between quasi-localized low-frequency modes and structural defects in colloidal crystals is obvious in Fig.~\ref{fig:gb}c. In particular, such modes in crystals appear to single out structural defects susceptible to external perturbations such as dislocations or interstitials; we note that they are less effective at picking out vacancies, which are mechanically more stable. Note that the correlation between quasi-localized modes and structural defects is robust to variation of the participation ratio cutoff between 0.1 and 0.3~\cite{suppl}.

The significance of this result lies in the fact that dislocations are known to be the defects that dominate the plastic response of crystalline solids~\cite{gtaylor}.  The fact that quasi-localized modes pick out dislocations indicates that they also pick out regions of the sample that are known to be susceptible to rearrangements.  Therefore, quasi-localized modes are concentrated in regions prone to rearrangements not only in disordered solids~\cite{manning2011,chen2011}, but also in crystalline ones.  This suggests that such modes may be general identifiers of flow defects for systems spanning the entire gamut from the perfect crystal to the most highly disordered glass.  Higher-frequency modes may provide a different link between these two extremes; it has been suggested that the boson peak at somewhat higher frequencies~\cite{chumakov2011} is related to the transverse acoustic van Hove singularity of crystals.  Our observation of the van Hove singularity, combined with earlier observation of the boson peak~\cite{chen2010}, shows that the covariance matrix method can be used to establish whether this relation exists in colloidal systems with tunable amounts of order.

We thank T.~C. Lubensky, V. Markel, P. Yunker and M. Gratale for helpful discussions. This work was funded by DMR12-05463, PENN-MRSEC DMR11-20901, NASA NNX08AO0G, DMR-1206231 (K.B.A.), T.S. acknowledges support from DAAD. 

\end{document}